# Theory of magnetism with temporal disorder applied to magnetically doped ZnO

Gillian A Gehring,[1]; Mahrous R Ahmed[1,2], Alex J. Crombie [1]

1. Department of Physics and Astronomy
   University of Sheffield
   Sheffield S3 7RH
   UK
2. Physics Department,
    Faculty of Science,
    University of Sohag,
    Sohag, Egypt.

**Abstract**

A dynamic model of the asymmetric Ising glass is presented: an Ising model with antiferromagnet bonds, $-J$, and ferromagnetic bonds, $+J$, with probabilities $q$ and $1-q$. The dynamics is introduced by changing the arrangement of the antiferromagnetic bonds after $n$ Monte Carlo steps but keeping the same value of $q$ and spin configuration. In the region where there is a second order transition between the ferromagnetic and paramagnetic states at $T_c(q)$ for $q < q_0$ the dynamic behaviour follows that expected for motional narrowing and reverts to the static behaviour only for large $n$. There is a different dynamic behaviour for $q > q_0$ where there is a first order transition between the ferromagnetic and spin glass states where it shows no effects of motional narrowing. The implications of this are discussed. This model is devised to explain the properties of doped ZnO where the magnetisation is reduced when the exchange interactions change with time.

**I Introduction**

The Ising model contains no intrinsic dynamics but has very interesting time dependent properties in models where dynamics has been added. These models include the Glauber model[1,2] in which a transition probability is introduced that allows the spins to relax to equilibrium and the Kawasaki model[3] that allows equilibrium to be established through spin exchange. A more relevant dynamic model for our problem is that where the spins are subject to a magnetic field that fluctuates randomly from positive to negative values distributed between $\pm h_0$; in this case a phase diagram is mapped out in $h_0-T$ space[4,5]. At $T=0$ a sufficiently large random field destabilises the uniform magnetisation but the field required to do this falls to zero at the critical point.

The Ising model is also a particularly interesting starting point from which to study frustration and there has been great interest in the Ising spin glass where the exchange interactions between spins are bimodal random variables[6]. A model in which the interactions have unequal probability of being ferromagnetic and antiferromagnetic has also been studied extensively. In this case the ferromagnetic, paramagnetic and spin glass phases meet at a multicritical point on the Nishimori line[7,8]. It has been found that the spin relaxation times diverge as the transition approaches[8]; this occurs because there is a multiplicity of states that are either degenerate with the ground state or are separated by a very low energy but nevertheless have very different configurations.

Recently an interesting physical phenomenon has led us to consider a new dynamic version of the Ising model where the exchange interactions fluctuate in time between $\pm J$ according to a probability distribution. The time is represented here by the number of Monte Carlo sweeps that are performed between the redistribution of the $J$ values. We explain the reasoning that has led us to develop this model and then describe the detailed model and the results of our simulations.

**II The motivation from the study of a real physical system**

In magnetically doped insulating oxides, particularly ZnO, there are free spins associated with transition metal (TM) ions, such as manganese $Mn^{2+}$ or $Co^{2+}$ that have been substituted on to the Zn sites at a concentration of typically ~5%. These are coupled together in two different ways. First there is the well known super-exchange through the intervening oxygen. This is always antiferromagnetic and is responsible for the antiferromagnetism that occurs in MnO and CoO.

However there is anther interaction that is special to doped semiconductors that is known as coupling by magnetic polarons[9]. This model describes insulating ZnO and other oxides that are *n*-type as grown and so must contain a number of donor sites – these may be oxygen vacancies or Zn interstitials or they may be added deliberately by doping with a trivalent impurity such as aluminium. The radii of the donor orbits are estimated to be larger than that for a hydrogen atom by a factor of $\varepsilon/m^*$ since the dielectric constant, $\varepsilon$, is large ~7 and the effective mass in the conduction band, $m^*$, is less than the free electron mass by a factor of ~0.3. The radius of these donor orbits can be as large, typically ~ 10Å, which is large compared with a lattice spacing, hence one donor state will overlap many transition metal ions. There is an exchange interaction between the electron in the donor orbit and the localised spins, the electron in the localised orbit becomes polarised, and hence it is called a polaron. Where the spins of two TM ions are coupled to one polaron these interactions give rise to a *ferromagnetic* interaction between the two localised spins. In this model the ferromagnetic transition temperature varies linearly with the number of polarons. This agrees with experiment as the observed magnetisation depends on the presence of defects[10]. Ferromagnetism occurs where the ferromagnetic interactions between TM ions due to the interaction with the polarons dominates over the antiferromagnetic superexchange interaction that is always present. Large moments have been observed in samples that are strongly insulating[11] so although there are a large number of polaronic states they do not conduct electricity.

It has been shown recently that the magnetisation drops sharply as the conductivity rises[12]. Mott variable range hopping,VRH,[13] occurs when the density of defect states is high. Hence we have a situation in which magnetism is favored by having a large density of defect states but is disfavored if the high density of defect states allows the electrons to become mobile. It is conjectured that magnetism is reduced because the motion of the electrons produces a time dependent exchange interaction.

When an electron moves from an occupied defect level, magnetic polaronic state, to an unoccupied defect state the exchange interactions between the TM ions in the vicinity change abruptly. Before the electron moved there would be ions that are coupled by a ferromagnetic interaction through the polaron and others at lattice sites where there was no polaron that had an antiferromagnetic exchange interaction due to superexchange[9]. After the polaron moves the exchange interactions between some of these ions may change from ferromagnetic to antiferromagnetic and vice versa. Hence electrical conduction is associated with *time dependent* exchange interactions between the TM ion spins.  In doped ZnO the relevant time scale is the time that a given donor state is occupied because this is the length of time over which the exchange is constant.  There are two relevant time scales for a magnet.  There is the local spin precession time (Larmor frequency) and the thermal relaxation time.  In this paper we consider the effects on magnetism due to the exchange varying on a time scale that is characteristic of the thermal spin relaxation time.  This is done using an Ising model where the exchange fluctuates between $\pm J$ as a function of time.

**III The model**

We consider a model of spins, $\sigma_n = \pm 1$, on a two dimensional square lattice and a three dimensional simple cubic lattice interacting with nearest neighbors with the probability of antiferromagnetic and ferromagnetic bonds given by $q$ and $1-q$ respectively;

$$H = -\sum_{<nm>} J_{nm} \sigma_n \sigma_m \qquad [1]$$

$$P(J_{nm}) = (1-q)\delta(J_{nm} - J) + q\delta(J_{nm} + J). \qquad [2]$$

At $T=0$ there is a critical value of $q=q_0$ such that the system is ferromagnetic for $q<q_0$ and has a spin glass phase for $q>q_0$. In one dimension the ferromagnetic ground state is destroyed for any $q \neq 0$ and in two dimensions $q_0$ was found by simulation[14] to be ~0.16,  In three dimensions it is known that there is a first order phase transition at $q_0$=0.23180(4)[7,8] which occurs below temperature $T=T^*$ where $T^*$=1.669$J$[8]. For temperatures below the critical temperature for the Ising model, $T_{Ising}$=4.5115$J$ but such that $T^*<T<T_{Ising}$ there is a second order transition between the ferromagnetic and paramagnetic phases as a function of the fraction of antiferromagnetic bonds, $q$. For temperatures below $T^*$ there is a first order transition between the ferromagnetic phase and the spin glass phase. The phase diagram that we obtained by simulation is shown in Fig. 1 it agrees with the published results[8].

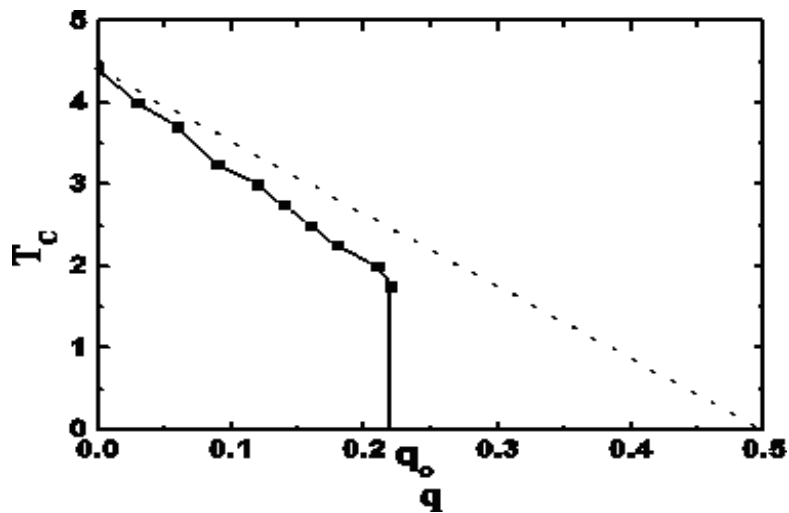

Fig.1 Plots of $T_c(q)$ found from the converged simulation, $n=10^5$, solid line, and from the motionally narrowed theory equan. 3b, dotted line.

We consider a model in which the exchange interactions are changed after a given time where the reallocated values are chosen according to the probabilities in equan. (2). The characteristic time between changes of the exchange interactions, $\tau$, should be compared with time taken for the spins to reach equilibrium, $\tau_0$. If $\tau \gg \tau_0$ then we should recover the static limit[7,8]. It is interesting to consider the motionally narrowed limit that occurs when $\tau \ll \tau_0$ where the distribution over the exchange constants may be replaced by the average,

$$<J_{nm}> = (1-q)J - Jq = J(1-2q) \qquad [3a]$$

$$T_c(q) = T_{Ising}(1-2q) \qquad [3b]$$

In this limit the spin glass phase does not occur and the ferromagnetism persists up to $q=0.5$ ($>q_0$) as shown by the dotted line on Fig. 1. When the motional narrowing theory is valid the magnetisation will *increase* for short $\tau$. The value of $\tau_0$ depends on the distance from the critical line and so is a function of both ($q_0-q$) and the temperature. In the Monte Carlo simulation the value of $\tau_0$ is represented by $n_0$ – the number of Monte Carlo sweeps required to obtain equilibrium.

**IV The Method and Results**

We use a Monte Carlo simulation with the Metropolis algorithm. After a chosen number of iterations, *n*, we note all the spin values and then all the exchange interaction are re-evaluated according to the probability distribution given in equation [2] and the simulation restarted with the same spin values and run for another *n* cycles. The procedure was repeated until $10^5$ Monte Carlo cycles had been performed. The magnetisation is evaluated as an average over the values obtained after the *n* Monte Carlo cycles.

Our first work was on a square lattice $L \times L$ where $L = 100$[13] but the results presented here were performed on a simple cubic lattice of size $L \times L \times L$ where $L = 20$. The results of the simulations are presented as a function of *q* at fixed temperature. We know from the phase diagram[8] that for $T>T^* =1.669J$ there is a second order transition between the ferromagnetic and paramagnetic phases as a function *q*. In Fig. 2 we show the results of the evaluation of the magnetisation from the simulations for *n* between *n*=1 and *n*=$10^4$ at T=3*J* and *T*=1.5*J* as these represent temperatures which are well above and just below *T\**.

We compare the values of the magnetisation with those found at equilibrium which corresponds to a value of *n*=$10^5$. The simulations done for *n* =1 show that the magnetisation is lower than for *n* =10 or 100 in both cases. This corresponds to the case in which the exchange interactions are being changed every Monte Carlo sweep and the full magnetisation cannot be established.

It is clear from Fig. 2 that the results of the simulation for $n>1$ are qualitatively different for $T=3J$ where the transition is second order compared with $T=1.5J$ where there is a first order transition to the spin glass phase in the static limit.

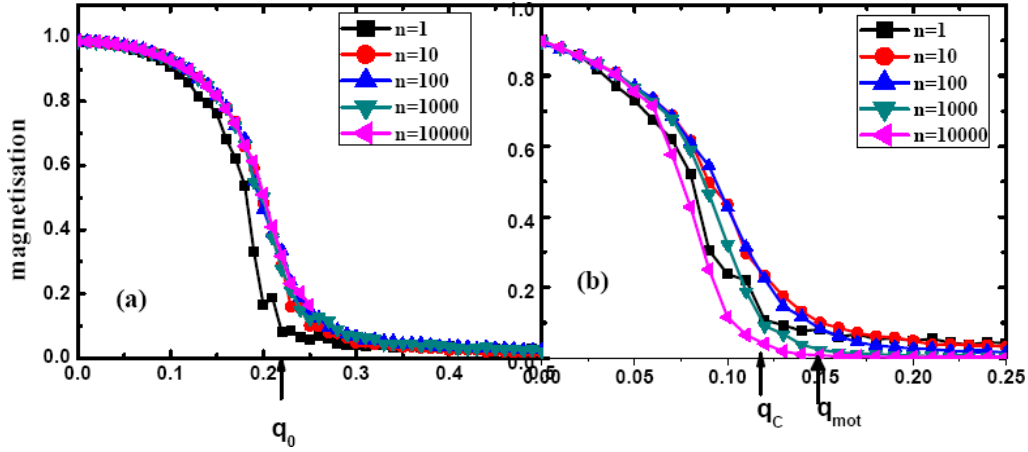

Fig. 2 (Colored on line) The variation of the magnetisation as a function of $q$ for fixed temperature at (a) $T=1.5J$ and (b) $T=3J$. In Fig 2a the critical value, $q_0$ is shown and the converged value of $q_c$ and the value obtained from motional narrowing $q_{mot}$ are shown in Fig 2b.

The theory of motional narrowing describes the results for $T=3J$; the critical value of $q$ is ~0.12 in the static limit and ~0.16 for full motional narrowing, given by equan [3b] as shown in Fig 1. The simulations for $n=10$ and $n=100$ agree with the theory obtained assuming full motional narrowing. For $n>100$ the results are between the motionally narrowed result and the static limit. This means we have established that the spins are thermalising over $n_0$~ 100 iterations in this case.

The results of the simulations for $T<T^*$ show that relaxation to the glass state is extremely efficient so that the static result is obtained for all $n>1$. In this case there are sufficient antiferromagnetic bonds to reverse the directions of some spins and the motional narrowing theory does not apply.

**Conclusion**

We have found that temporally fluctuating exchange affects the magnetisation differently depending on whether the transition is first or second order. The motionally narrowed theory is valid in the second order region – in the case shown here $T=3J$. The motional narrowed result was found for $n<100$ indicating that the number of Monte Carlo cycles needed for stabilisation is $>100$. A different result is found when there is a first order transition to the glass phase; in this case the magnetisation is essentially independent of $n$ for all $n >1$. This can be understood when it is realised that in the Monte Carlo method a spin will always flip if it is energetically favourable to do so. The large number of antiferromagnetic bonds in the spin glass phase means that these flips occur at the transition to the spin glass phase. If it is not energetically favourable to flip a spin then the Monte Carlo method will still allow it to flip with a certain probability depending on the temperature. A small number of antiferromagnetic bonds will have the effect of weakening the exchange field at any given site and it is in this range that the motional narrowing theory holds. The effects calculated here do not explain the results observed in doped ZnO. We note that in a real spin system there are two relevant relaxation times, the thermal relaxation time that is relevant for the Ising model and the Larmor precession time that occurs for dynamic spins. This study shows that allowing the exchange to vary over times comparable with the relaxation time will not give the desired result and so work on a dynamic model that considers the effect of a vector fluctuation is underway. The calculations presented here point to the richness of this dynamic version of the frustrated Ising model.

**Acknowledgments**

We should like to thank Professor Arghya Taraphder for very useful discussions about the model proposed in this work and Professor Thomas Schrefl for considerable help to AJC. We also thank the EPSRC for financial support.

**References**

1. Roy J. Glauber, J. Math. Phys. **4** 294 (1963).

2. Mauo Suzuki and R.Kubo, J. Phys. Soc Jap. **24**, 51 (1968).

3. K. Kawasaki, Phys. Rev. **145**, 224 (1966).

4. M. Achayya, Phys. Rev. E **58**, 174 (1998).

5. Bikas K. Chakrabarti and M. Achayya, Rev. Mod. Phys. **71**, 847 (1999).

6. K.H. Fischer and J.A. Hertz, *Spin Glasses*, Cambridge University Press (1991).

7. H. Nishimori, Prog. Theo. Phys. **66**, 1169, (1999).

8. Martin Hasenbusch, Francesco Parisen Toldin, Andrea Pelissetto, Ettore Vicari, , *Phys Rev B*, **76**, 094402 (2007).

9. J.M.D. Coey, M. Venkatesan, and C.B. Fitzgerald, *Nat. Mater.* **4**, 173 (2005)

10. F. Pan, C. Song, X.J. Liu, Y. C. Yang, and F. Zeng, Materials Science and Engineering **R62**, 1 (2008).

11. C Song et al *Phys. Rev. B*, **73**, 024405 (2006)

12. A.J. Behan, A. Mokhtari, H.J. Blythe, D. Score, X-H. Xu, J.R. Neal, A.M. Fox and G.A. Gehring, Phys. Rev Lett **100**, 047206 (2008).

13. N.F. Mott, *Metal Insulator Transitions* (Taylor and Francis 1990).

14. A. J Crombie *MPhys thesis* University of Sheffield (2008).